# Making of a nonlinear optical cavity
## Cómo construir una cavidad óptica no lineal

R. Martínez-Lorente,[1*] G.J. de Valcárcel,[1] A. Esteban-Martín,[2] J. García-Monreal,[1] E. Roldán,[1] and F. Silva[1].

1. *Departament d' Optica, Universitat de València, Dr. Moliner 50, 46100-Burjassot, Spain*
2. *ICFO Institut de Ciències Fotòniques, Mediterranean Technology Park, 08860 Castelldefels, Spain*

**ABSTRACT:**

In the article we explain in detail how to build a photorefractive oscillator (PRO), which is a laser-pumped nonlinear optical cavity containing a photorefractive crystal. The specific PRO whose construction we describe systematically, is based on a Fabry-Perot optical cavity working in a non-degenerate four wave-mixing configuration. This particular PRO has the property that the generated beam exhibits laser-like phase invariance and, as an application, we show how a suitably modulated injected beam converts the output field from phase-invariant into phase-bistable. While the emphasis is made on the making of the experimental device and on the way measurements are implemented, some introduction to the photorefractive effect as well as to the necessary concepts of nonlinear dynamics are also given, so that the article is reasonably self-contained.

**RESUMEN:**

En este trabajo explicamos detalladamente cómo construir un oscilador fotorrefractivo (PRO), que es una cavidad óptica bombeada con láser que contiene un cristal fotorrefractivo. El PRO concreto cuya construcción es descrita paso a paso, está basado en una cavidad óptica lineal (tipo Fabry- Perot) funcionando en una configuración de mezcla no degenerada de cuatro ondas. Este PRO particular tiene la propiedad de que el haz generado exhibe emisión laser con invariancia de fase y, como aplicación, mostramos cómo un haz inyectado y modulado adecuadamente convierte el campo de salida con invariancia de fase en uno con biestabilidad de fase. Aunque se ha hecho énfasis en cómo configurar el dispositivo experimental y en cómo realizar las medidas, se incluye una introducción al efecto fotorrefractivo así como a los conceptos necesarios de dinámica no lineal, resultando un trabajo razonablemente auto contenido.

**Palabras clave:** Efecto fotorrefractivo, mezcla de ondas, óptica no lineal, resonador óptico, patrones, biestabilidad.

# 1. Introduction

Experimental research articles seldom contain many details about the construction and tuning of the experimental device. As this essential knowledge is considered, somehow, as a sort of *common knowledge*, an inexperienced researcher willing to implement in the lab an experimental device will hardly find enough details in the research literature, these details being actually transmitted personally to the beginner by their more experienced colleagues. Certainly, building and tuning details do not use to involve new knowledge, which justifies passing quickly through them in published reports in order to concentrate on the specific goal of the research. However, it would be useful, especially for researchers who are novel in a particular experimental field, to have at hand primers where at least some of "the things that are never explained" could be found. This could also be inspiring for the design of experiments aimed at teaching in undergraduate levels. This is the context of the article in which: (i) we explain in *full* detail how to build a particular optical device, namely a photorefractive oscillator, and (ii) how to make experiments on nonlinear pattern formation with it.

A photorefractive oscillator (PRO) [1] is a special type of nonlinear optical cavity [2, 3], a term used to generically refer to optical resonators that contain a nonlinear medium (a material exhibiting some kind of nonlinear optical response); in the case of the PRO the nonlinear medium being a photorefractive crystal (PRC) [6, 7]. Some PROs find applications in phase conjugation or gyroscope, among many others [6, 7]. However, our interest is not on the applicability of the device: We are interested in the study of the PRO as a nonlinear optical cavity, i.e., we use it as an *experimental model* of such cavities. One of their characteristics that is more appealing for us is the usually slow response time of PRCs (ranging from tenths of $\mu s$ to tenths of $s$), which has the advantage that no fast electronics is necessary for the observation/ recording of the dynamics of the system output. This simplifies the experimental resources needed for research in nonlinear patterns as compared to other nonlinear optical cavities containing faster material media such as, e.g., semiconductors.

As stated, for us PROs are just a useful model for the study of some aspects of nonlinear optical cavities. More specifically, it turns out that from a nonlinear dynamics perspective, the type of temporal dynamic regimes, as well as the type of patterns displayed by a particular nonlinear optical cavity, depend only weakly on the particular type of nonlinearity, being *universal* [8, 9] many of the features of the dynamics. In particular, the symmetry properties of the phase of the intracavity field determine many features of the spatiotemporal dynamics of the system. PROs are ideal systems for studying this phase properties because depending on the type of cavity and on the way of pumping it, they can display very different phase properties.

We shall comment more below about that physics, but this exciting research *is not* the main goal of this article, but the making of the device itself. If we think that describing the building of the setup could be of interest to others is because, in essence, we are just mounting an optical cavity and running an optical experiment. This means aligning mirrors and lenses, controlling polarizations, injecting laser beams, making interferograms, stabilizing the cavity length, etc., activities that are common to most optical laboratories irrespective from the specific research carried out in them.

After stating the aim of the article, the rest of it organizes as follows. In Section II, we briefly review the physics of the photorefractive effect, and in Section III we make a general description of the different types of PROs depending on the type of cavity (ring vs. linear) and the modality of pumping (unidirectional vs. bidirectional). We pay attention to the main feature we are interested in, namely the phase properties of the output field (phase invariant vs. phase bistable). Section IV is the most important and we describe in it, systematically, how to build a Fabry-Perot-type cavity PRO with plane mirrors and unidirectional pumping. In this case, the intracavity nonlinear interaction is a non-degenerate four–wave mixing process, which leaves the output field phase invariant, as it occurs in a free running laser. As will be explained below, an important requisite for our device is that its cavity length could be precisely tuned, which we solve with a double-cavity technique and an active stabilization mechanism. Moreover, as we are interested in having a very short effective length in the cavity, because we want to reach a very large Fresnel number, the cavity we build has telescopes within, so that their tuning allows fine variations of the effective cavity length. This is a special type of so-called self– imaging resonators [10], first used as we do it by Carl O. Weiss and coworkers [11]. These telescopes, as a by-product, permit an easy access inside the cavity to the far field distribution, so that filtering is easy. Once the device construction and tuning have been fully described in Section V, we pass, in section VI, to concentrate on a particular experiment that consists in transmuting the system from phase invariant to phase bistable, which is accomplished by injecting in the PRO an additional laser beam having amplitude modulation (a technique dubbed *temporal rocking*). Then in Section VII, we give our main conclusions and outlook.

## 2. The photorefractive effect

The photorefractive effect is a nonlinear phenomenon that occurs in some non- centrosymmetric crystals, and consists in the appearance of a light-intensity dependent spatial modulation in the refractive index of the material when submitted to a *spatially varying* illumination. The phenomenon was first observed in 1966 by Arthur Ashkin and coworkers at Bell Laboratories when experimenting with the inorganic crystals $LiNbO_3$ and $LiTaO_3$ [12], and has found quite many applications, especially in holography and phase conjugation [4, 5].

Even if it brings some similarity with the optical Kerr effect, the photorefractive response is quite different and more complex because charge transport processes and interband transitions are involved in it [4–7]. Let us be somewhat more concrete. In the Kerr effect the refraction index has the form $n = n_0 + n_2 I$ ($n_0$ is the linear refraction index, $I$ is the light intensity, and $n_2$ is the nonlinear index that can be positive or negative), a phenomenon that appears in centrosymmetric nonlinear media [5]. The photorefractive effect, contrarily, appears in non-centrosymmetric media. As it also manifests as a dependence of the refraction index on the incident fields (a more complicated one than that of the Kerr effect, see below), two- and four-wave nonlinear mixing can occur in photorefractive media similarly as they do in Kerr media.

The photorefractive nonlinear response can be understood as the combined result of the electro-optic Pockels effect (i.e., the dependence of the refractive index on the amplitude of an applied electrostatic, or quasi-electrostatic, electric field, a nonlinear phenomenon that appears in non-centrosymmetric crystals) together with the appearance, within the crystal, of a so–called *spatial charge field*, $E_{sc}$ [4, 5]. The field $E_{sc}$ originates within the crystal in the diffusion of the charges that have been excited by the light field, which move from illuminated to non–illuminated regions, see Fig. 1. (In some cases, an extra electric field needs to be added to the PRC with the purpose of helping diffusion in moving the charges.) Then, the inhomogeneous light distribution creates a spatial separation of charges that move escaping from illuminated regions, which creates an $E_{sc}$ within the crystal that, because of the Pockels effect, causes a change in the refractive index.

The critical feature is that the index grating is dephased with respect to the light intensity grating, which makes possible the transfer of energy from, e.g., one of two interfering light beams to the other one [4, 5]. Let us be a little more specific. Consider a photorefractive medium illuminated by the light field

$$\boldsymbol{E}(\boldsymbol{r},t) = \boldsymbol{e} \sum_{j=1,2} A_i(z) e^{+i(\boldsymbol{k}_j \boldsymbol{r} - \omega_j t)} \quad (1)$$

with **e** a unitary polarization vector, $A_j(z)$ the slowly varying amplitude of the beams propagating along the photorefractive crystal; $k_j$ and $\omega_j$ are the wavevectors and frequency of the amplitude components respectively. In this medium an index grating forms

$$\Delta\left(\frac{1}{n^2}\right) = r_{ijk} E_{sc,k} \quad (2)$$

where $r_{ijk}$ are the components of the electro-optic tensor (summation over $k=x,y,z$ is assumed), and $E_{sc,k}$ is the k-th component of the space–charge field inside the medium, which for diffusive type photorefractive crystals (i.e., without additional electric bias) is given by [4]

$$\boldsymbol{E}_{sc} = Re[E_{sc} e^{-i\boldsymbol{K}\cdot\boldsymbol{r}}]\boldsymbol{u}_K \quad (3a)$$

$$E_{sc} = i \frac{E_d}{E_d + E_q} \frac{A_1(z) A_2^*(z)}{|A_1(z)|^2 + |A_2(z)|^2} \quad (3b)$$

with $\boldsymbol{u}_K$ a unitary along the intensity grating vector $\boldsymbol{K} = \boldsymbol{k}_1 - \boldsymbol{k}_2$. $E_d$ and $E_q$ above are the so-called diffusion and saturation fields, which are characteristic of the particular photorefractive crystal. The above expression reveals the dependence on the complex field amplitudes, and the "*i*" factor in $E_{sc}$ reflects the dephasing of the intensity grating with respective to the refractive index grating, by π/2.

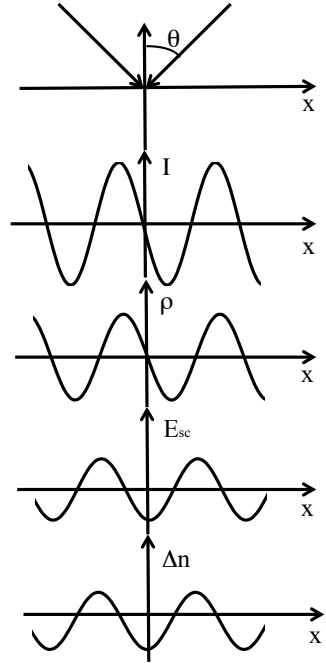

Fig.1. Photorefractive effect. Two beams interfere with angle θ, creating a harmonic intensity distribution *I(x)*. Hence, there are bright regions that alternate with dark ones, and charge mobility favors charge accumulation in dark regions so that a spatial charge distribution *ρ* appears. The electro-optic effect translates the space charge field $E_{sc}$ into a spatial variation of the refractive index, the index grating *Δn* being dephased π/2 with respect to the intensity grating. The dephasing is necessary for the transfer of energy from one beam to the other.

In short, when two light beams illuminate the diffusive PRC this is well characterized as having an effective refractive index $n_{eff}$ of the form

$$n_{eff} = n_0 + i\frac{n_1}{2}\frac{A_1 A_2^*}{|A_1|^2+|A_2|^2}e^{i\boldsymbol{K}\cdot\boldsymbol{r}} + c.c \qquad (4)$$

with $n_0$ the refractive index in the absence of illumination and $n_1$ proportional to the space-charge field $E_{sc}$. When more than two light beams interfere, as in four wave mixing (FWM) processes, more index gratings must be taken into account, but two-field gratings usually dominate the interaction; however we do not need to go further in this direction and refer the interested reader to the literature for further knowledge [4-7].

The modulation of the index grating is peaked for particular polarizations of the interacting fields and for a particular orientation of the grating vector ***K*** with respect to the symmetry axis of the crystal, its c-axis. In our case the PRC is a 4.5x4.5x8mm³ BaTiO₃ crystal, and the fields must be polarized on the plane containing the crystal c-axis, as well as the two beams must form an angle of δ=45° and α=8° with respect to the c-axis, see Fig. 2, where the configurations are shown for two- and four-wave mixing. There is of course a tolerance in the angular values, which can be rather high (several degrees).

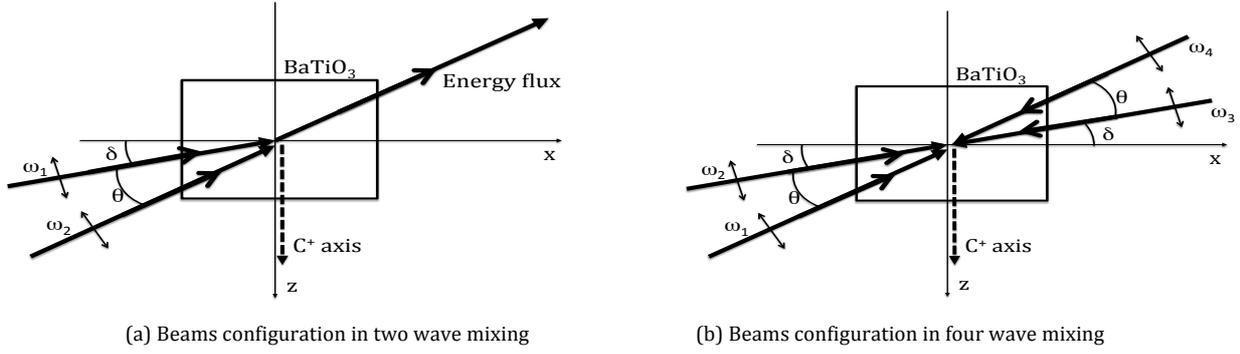

(a) Beams configuration in two wave mixing

(b) Beams configuration in four wave mixing

Fig.2. (a) In the two-wave mixing (TWM) configuration two beams with the same polarization impinge on one of the faces of the photorefractive crystal. A wide cone of emission appears within of a certain solid angle, a phenomenon called *beam fanning* (see 5.b.4). (b) In the four wave mixing (FWM) configuration, two pairs of beams, again with the same polarization, impinge on opposite faces of the crystal.

### 3. Classification of photorefractive oscillators

The classification can be made attending to the type of optical cavity: ring cavities, in which the cavity modes are traveling waves, and linear cavities (or Fabry-Perot type cavities) in which the cavity modes are standing waves. However, the truly relevant fact from the nonlinear dynamics viewpoint is whether the wave mixing occurring inside the resonator is degenerate or non-degenerate. In the former case the system can be phase invariant (i.e., the output field phase can take any value), while in the latter the system is phase bistable (the output field phase takes one among two fixed values differing by $\pi$).

Consider ring cavities first. If the PRC is pumped by a single pumping field (unidirectional pumping), see Fig. 3(a), there is a two-wave mixing process inside the cavity by virtue of which photons from the pump beam are efficiently transferred to the cavity mode. Of course, this occurs for particular pumping and signal modes directions, which forces the signal beam to be in only one of the two possible intracavity counter-propagating traveling wave modes. This two–wave mixing process leaves the signal mode (the mode amplified within the cavity) phase invariant because its generation does not depend on any particular phase relation with respect to the pumping field; they mix at the *intensity* grating. Hence, this system is similar to a laser in which the amplification process leaves the phase of the field undetermined (this is not the only similarity existing between lasers and two-wave mixing PROs [13]).

If the crystal is bi-directionally pumped, see Fig. 3(b), it is a four-wave mixing (FWM) process that occurs, because the ring cavity can sustain bidirectional oscillation. Again, the phases of the generated fields (the two counter-propagating intracavity signal fields) are not fixed by the interaction, because the two counter-propagating waves do not need to have any spatial relation among them. The process is again phase invariant.

In contrast with ring cavities, in linear cavities two-wave mixing is impossible as the cavity imposes that the signal field be necessarily bidirectional (it is a standing wave), hence the nonlinear process is FWM. There are two possibilities: that the cavity be pumped by a single pumping beam, as in Fig. 3(c), or by two counter-propagating pumping beams, Fig. 3(d). It turns out that in the former the interaction is non-degenerate while it is degenerate in the latter. The degeneracy of the latter case occurs because the phases of two of the waves (the pumping beams) are fixed, and the linear cavity locks the phases of the two counter-propagating signal modes. Indeed the phase of the signal field is locked to *two* possible values that differ by $\pi$, hence the system is said to be *phase–bistable*.

In resume, ring cavities permit non-degenerate TWM (or FWM) when one (or two) pump(s) are used; while linear cavities permit non- degenerate (or degenerate) FWM when one (or two) pump(s) are used. Again, we refer the reader to the literature for details. In particular, in [4] the coupled fields equations for the four processes just mentioned are given.

### 4. The goal of the experiment: converting a phase-invariant oscillator into a phase bistable one.

As already advanced, from our research perspective the important feature of a photorefractive cavity is the phase invariance, or not, of the signal field. We pass now to comment briefly, which are the differences between a PRO with phase invariance and a phase bistable one.

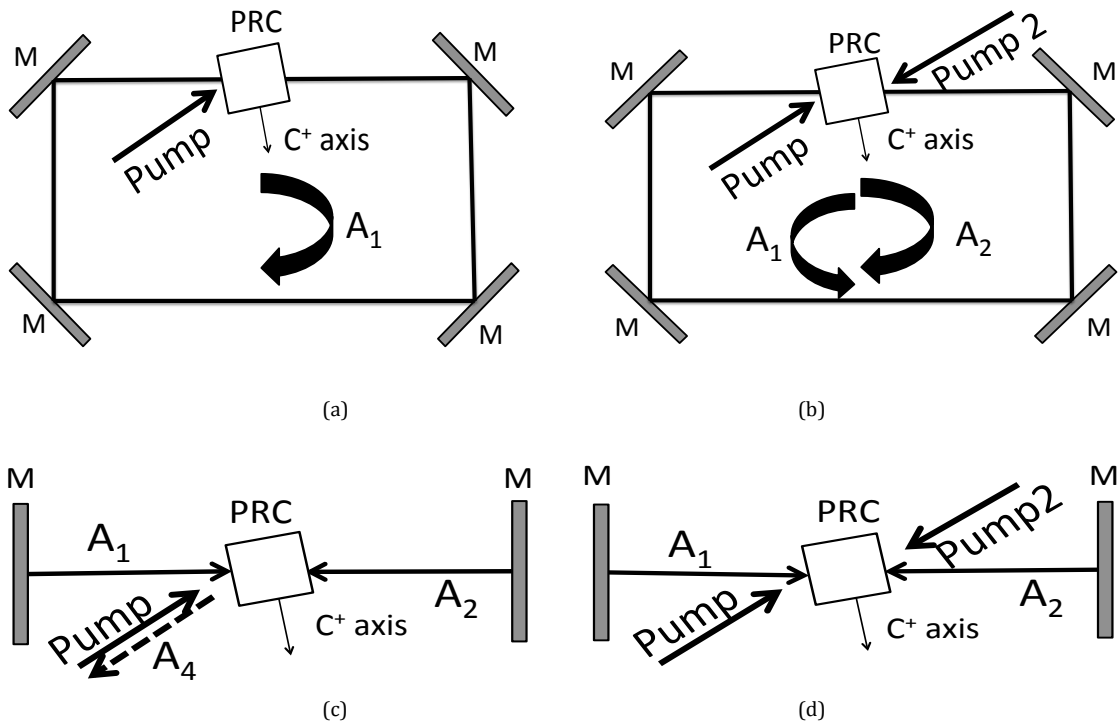

Fig.3. Implementation of different nonlinear wave mixing processes in photorefractive oscillators. Figures (a) and (b) show ring cavities differing in the way they are pumped: either singly (a) or with two counter-propagating pump beams (b). The corresponding nonlinear processes are nondegenerate TWM and nondegenerate FWM. Figures (c) and (d) show a linear cavity with one (c) and two (d) pumping beams, the corresponding processes being degenerate and non-degenerate FWM.

Consider first a single mode PRO, i.e., the intracavity field is on just one of the resonator modes. (This depends on the resonator geometry, but is easy to control with an intracavity iris). What one would observe is that when the PRO is phase invariant the phase of this mode diffuses with time, while it remains locked to the pump field phase when it is not phase invariant. The phase locking can be mono-stable (as it occurs, e.g., in a laser with injected signal, in which the laser phase locks to that of the injection), or can multi-stable, in particular bistable, as we have commented that occurs in the degenerate FWM process. Then, when the system is phase *bistable*, in different runs one would observe that the signal phase takes the value $\phi$, or the value $\phi + \pi$ in different runs. The difference is much more dramatic when extended patterns form in multimode resonators, as the type of patterns that form are completely different.

About patterns [2, 3, 8, 9], when the number of transverse modes sustained by the oscillator is very small, the complexity of the intensity distribution (in the plane transverse to the light propagation direction) is small too. But when there is a very large number of modes, or better a continuum of modes, the intensity distribution can display extended patterns [2, 3] that belong to the general category of dissipative structures [8, 9]. These patterns form in the plane transverse to the cavity axis, and manifest in the modulation of the intensity distribution, that can be periodic in space (hexagons, stripes, squares), or aperiodic (labyrinths), or can consist in a number of localized structures. Of course, the patterns can also change with time and the resulting spatiotemporal complexity can be very rich. Interestingly, when the boundary conditions do not impose a particular symmetry (as it occurs in a plane–mirror cavity), the mathematical models that describe all these phenomena connect with analogous equations found in very different fields (chemistry, biology, other physical systems) [8, 9]. This is why pattern formation is frequently said to be a universal phenomenon, in the sense that the mechanisms that lead to the appearance and selection of the patters are very general and weakly dependent on the particular system.

A particularly appealing type of nonlinear pattern are *localized structures*, such as vortices, cavity solitons, domain walls, etc. These type of structures consist in intensity variations that are local and that are independent from other distant localized structures. When they can be excited/erased without affecting neighboring structures, they are known as cavity solitons, which have potential applications for information processing [14–16].

The difference between systems with different phase symmetry is that in phase invariant systems vortices form spontaneously, while in phase bistable systems domain walls form spontaneously. Vortices and domain walls are two very different types of localized structures see Fig. 4. In a vortex, the phase of the field accumulates 2π when circulating around the vortex core, which makes necessary the existence of a zero (a singularity) in the field intensity, then a vortex appears as a dark spot. Of course, the system must be phase invariant for exhibiting vortices because all phase values must be permitted. Vortices were first predicted to occur in nonlinear optical cavities in 1989 by Pierre Coullet and coworkers [17].

Contrarily, when only two (opposite) phase values are possible, localized structures are the borders of domains with different phase, these borders being called domain walls, which were first studied in the context of ferromagnetism.

The domain walls have chirality as the π phase accumulation across them can occur by increasing or by decreasing the phase. In 2D, the domain walls can close on themselves, hence separating an inner region with a phase from the surroundings having opposite phase, the core of the wall appearing as a dark circular line. These structures are known as ring dark-solitons [18].

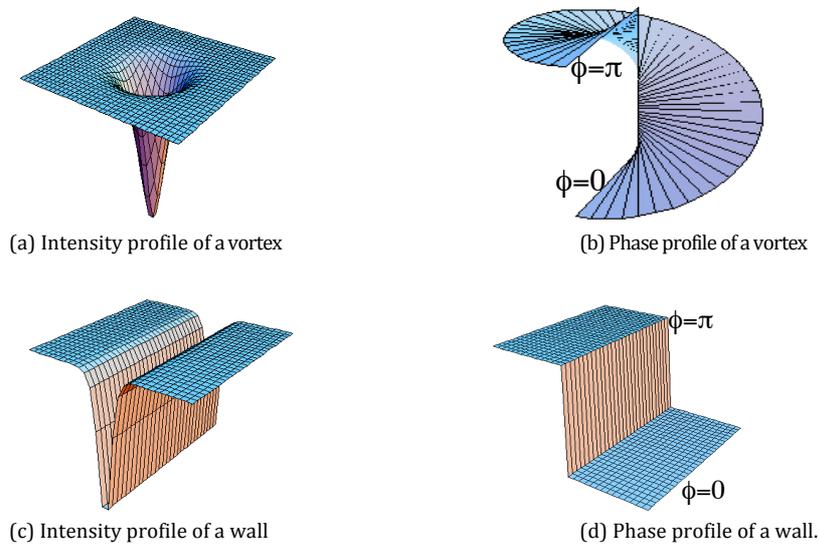

(a) Intensity profile of a vortex  (b) Phase profile of a vortex

(c) Intensity profile of a wall  (d) Phase profile of a wall.

Fig.4. A vortex as appears (top) is a complex structure in which the intensity is homogeneous but at the singularity, where it drops to zero (a), while its phase accumulates 2π around a singularity at which all phase values are permitted (b). The 1D domain wall (bottom) is a real structure in which there are two homogeneous states in intensity (c) and phase (d), but with an opposite phase value (differing by π). The intensity drops to zero at the boundary between the two opposite states.

For a nonlinear optical cavity can form extended patterns it is necessary that the cavity has a large Fresnel number. The Fresnel number is the angle subtended by one end cavity mirror as seen from the opposite end cavity mirror, and when it is large it means that a large number of transverse modes can oscillate (notice that an intracavity iris can easily change the Fresnel number). Moreover, in order to deal with an as ideal as possible system, it is highly recommended that the cavity has plane mirrors, as in this case there is a continuum of degenerate transverse modes having the same frequency.

The experiment we will describe consists in the demonstration of the so–called *rocking mechanism* [19-21]. Rocking consists in the injection on the system under study of a signal having (only) amplitude modulation at an appropriate frequency. The injected signal can be amplitude modulated in time (temporal rocking [19]) or in space (spatial rocking [22–25]); here we consider the first case. As we show below, rocking converts a phase invariant non–linear system into a phase bistable one and this is noticed by observing the type of transverse patterns the system forms: in the absence of rocking the system must exhibit vortices while when rocked it must exhibit domain walls

## 5. Building the nonlinear cavity

The PRO we build is a Fabry-Perot type cavity with plane mirrors and short effective cavity length in order to guarantee a large enough Fresnel number. In principle, one could think of making a very short cavity with large plane mirrors placed very close to the crystal facets. However, the length would still be limited to a minimum (the crystal length, at least) and the device would not be easy to handle because of lack of space. In spite of this, we build a large cavity (120 cm long in our case) that contains imaging forming systems (telescopes) inside the resonator that image the cavity mirrors on the crystal facets, or wherever. In fact, such a cavity can have a null effective length or even a negative effective length, and one of its advantages is that the far field plane is accessible inside the cavity, at the telescopes foci, which makes filtering particularly simple.

Next, we describe the main characteristics in the design of the cavity, namely the use of a double cavity for cavity length control, and the use of telescopes for making a self-imaging resonator. Later we describe the building and alignment of the system step by step.

### 5.a. Design of the cavity

The nonlinear optical cavity we are going to build must verify two important requisites for allowing the controlled observation of dissipative structures: it must have a large enough Fresnel number, and it must have a controlled length. The first requisite is essential for observing patterns, and the second is essential for keeping the patterns reasonably steady because some of the characteristics of the patterns (not all of them) depend strongly on detuning. These requisites are fulfilled by using a double cavity scheme (for detuning control) and by constructing a self-imaging resonator. The special design we present below is a variation of that implemented by Weiss in [11].

### 5.a.1. A double cavity for detuning control

Detuning, $\Delta$, is the difference between the frequency of the laser light beam pumping the cavity, say $\omega_p$, and the frequency of the cavity longitudinal mode closest to that frequency, say $\omega c$, so that $\Delta \equiv (\omega_p - \omega c)$. While the pumping frequency $\omega_p$ is fixed, the cavity frequency $\omega c$ depends on the cavity length, so that $\Delta$ changes by varying the cavity length. Hence, we need a cavity that maintains a fixed and controllable length. In order to have an intracavity signal that allows us to measure in real time its length variations, we play with the light polarization by using the double cavity shown in Fig. 5. The simple idea is the following: the cavity divides into two, of equal length, each of the cavities sustaining a different polarization (one horizontal, and the other vertical). While one of the polarizations interacts with the nonlinear crystal, the other does not, and is used only for measuring the cavity length, and controlling it.

The horizontal plane contains the cavity axis and the crystal c-axis so that for amplification it is necessary that the pumping beam has horizontal polarization. The device shown in Fig. 5 is injected with 45° linearly polarized light, and a polarization beam splitter (PBS) divides this pumping beam into two beams, one with horizontal (H) polarization will interact with the crystal, and one with vertical (V) polarization that follows the path in which there will be no crystal. The two beams are recombined into one with the help of two piezoelectric mirrors (PZMs) and a second PBS, see Fig. 5.

Hence, the two orthogonal polarizations follow different paths of equal length (except for the presence of the crystal in one of them, which is irrelevant for our purposes). As the V polarization plays no role in the interaction, we can use it at the output of the cavity for detecting variations (due to thermal and mechanical fluctuations) of the cavity length. These variations are detected, and corrected, with the help of a lock-in amplifier. Piezoelectric mirrors are used for changing the cavity length accordingly with any detected variation, so that the cavity length can be kept fixed through a feedback loop. Of course, the correction of the cavity length is made with some delay, but this is very short as compared to the characteristic time of the fluctuations, so that the correction is instantaneous for all practical purposes. Finally, in order to minimize errors, the two PBSs, the mirror and the crystal are all mounted on a rigid piece of invar, so that the part of the device shown in Fig. 5. is mounted apart, and later introduced in the resonator

### 5.a.2. Self-imaging resonator

If the two cavity end-mirrors were placed at the entrances of the device in Fig. 3(c), which has no telescopes, the cavity would have a too small Fresnel number unless the mirrors were absurdly large. In spite of this we use normal size plane mirrors (with diameter of 2 cm) and insert between each of them and the double– cavity device both telescopes that form the image of the end mirrors somewhere close the nonlinear crystal. In this way, an effective cavity length that can be varied over some range (including zero and negative lengths) is easily achievable. In our case we use an effective cavity length of 1-2 cm. The reason for using telescopes and not, e.g., single lenses, is that the image of the plane mirror must continue being a plane mirror; either the wavefronts would suffer some curvature. Moreover, in this case, the displacement of the image is linear with the displacement of the object, which is convenient for detuning control.

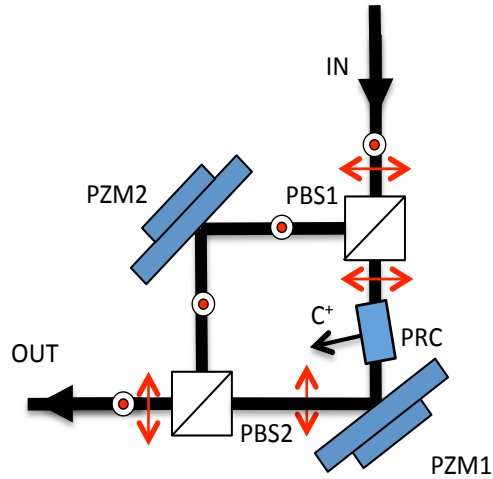

Fig.5. Double cavity arrangement for detuning control. Two co-linear beams enter the double cavity device with orthogonal polarizations (arrows mark horizontal polarization and circles mark vertical polarization). A first PBS (PBS1) separates each polarization that follows a different path. In one of the paths (horizontal polarization), the light interacts with the PRC, while in the other one there is not any nonlinear interaction. Finally, a second PBS (PBS2) recombines both beams into a single beam. Any variation of the optical path in the "active cavity" (horizontal polarization) is also suffered by the vertical polarization, which is used only for cavity length control.

As stated, in this device the far field of the cavity is placed at the common focal plane inside the telescopes. The near field is the field distribution one finds at the PRC facets, and the far field (the spatial Fourier transform of the near field distribution) is in principle detected far outside the cavity, but the telescopes approach this far field to the referred focal planes, see Fig. 6. This has the obvious advantage of allowing filtering inside the cavity.

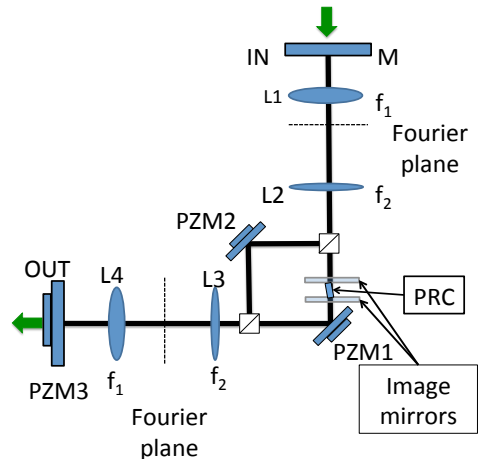

Fig.6. Self-imaging resonator. Two telescopic system image both end cavity mirrors on the faces of the PRC, so that a virtual cavity with the mirrors as close as one wants, can be made.

Hence two telescopes must be mounted, one in each of the arms of the cavity; we explain later the right moment for doing that. In our case, each telescope consists of two lenses with focal distances f = 10 cm and f = 20 cm separated, obviously, 30 cm (see Fig. 6), being the distance between the end mirror and its respective telescope 10 cm, and the distance between the closer lens of the telescopes and the PRC 20cm. The selection of focal lengths attends two goals, namely making an enlarged image of the plane mirrors on the PRC (they must be larger than the crystal), and not introducing large aberrations, which implies not using too short focal lengths.

## 5.b. Assembling the cavity

There is not a single way for mounting the device, of course, and the experience of the experimentalist is an important aspect at this stage. Next we describe what we think is a good procedure.

### 5.b.1. Extracting beams from the laser source

The first part to be mounted is that for extracting several laser beams from the single laser source. We must extract: the pumping beam, the cavity control beam (CCB in the following), which does not interact with the PRC, the reference beam for making interferograms of the cavity output, the rocking beam for making rocking, as well as an additional beam for injection purposes (see below). All beams have horizontal polarizations but the CCB, which has vertical polarization. A half-wave plate (HWP) is used to rotate the laser polarization (usually horizontal or vertical) at a convenient polarization angle. In our case it is such that the ratio of V to H polarizations powers is around 1/9. See Fig. 7 and its caption for more details.

As within the cavity the rocking beam has the same polarization and follows the same path as the mode that the cavity will generate, we will refer to them as cavity active beam in the following (CAB). In Fig. 7 we do represent how all these beams are extracted and distributed in space with the help of PBSs, PZMs and mirrors.

### 5.b.2. Assembling the cavity. Alignment

As stated, we first build and align the double cavity device of Fig. 5, then place the two end–mirrors, and then the telescopes. In fact, the detection mechanism must be mounted after the mirrors and before the telescopes, see below. Before positioning the elements on the worktable, distances must be taken in order to guarantee the disposal of enough room. The first end–mirror to be mounted is that farther form the laser source, and it is a PZM. It is highly advisable the use of two pinholes mounted on micrometer positioners (along both x and y) in order to improve the alignment by passing the laser beam trough them. We also use them (when necessary) as low frequency filters by positioning them on the Fourier planes, see below on detection and observation.

After the end–mirrors have been aligned, the first telescope to be mounted is, again, that farther from the laser. It is convenient that at least one of the two telescope lenses supports allow for micrometric displacement on the plane in order to achieve a good adjustment of the afocal condition.

There is a useful trick for getting good alignments and it consists in using two mirrors and two pinholes disposed as in Fig. 8. It allows that the beam can be displaced parallel to itself, and the pinholes help in testing the alignment. This is used with the rocking beam. The other beams are reflected by two PBSs (Fig. 7), the same can be done by slightly rotating the PBSs because the beam they transmit does not significantly deviate when the prism is rotated a very small angle.

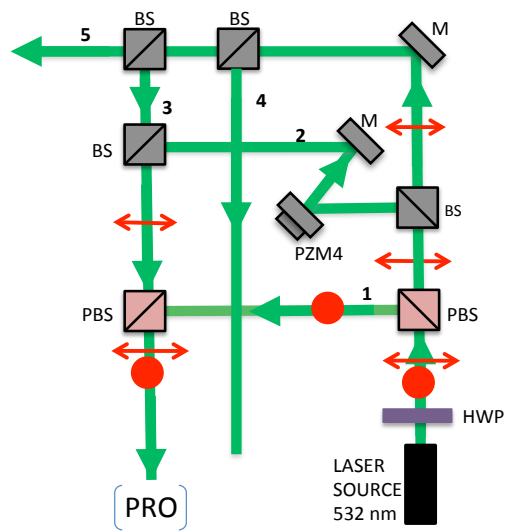

Fig.7. Schedule of beams preparation. We use beam splitters (BSs), PBSs and mirrors (M) in order to address each beam with a different proposal. The beam number one is the CCB, the beam number two is the rocking beam, the beam number three is the CAB, the beam number four is the pump beam and the beam number five is used as reference to make interference.

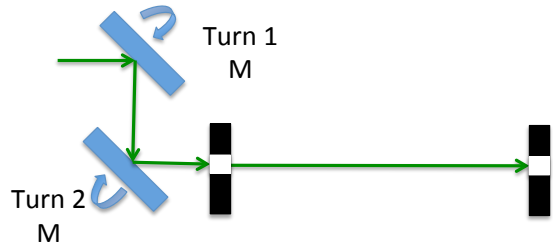

Fig.8. Experimental procedure for efficiently aligning a beam. Mirrors M1 and M2 can be rotated, so that the light beam can be displaced parallel to itself. Pinholes 1 and 2 help in checking the correctness of the alignment.

It could occur that the alignment of one of the beams (either the CCB or the CAB) implies some misalignment in the other beam. We choose the CAB as the master, and we use PZM2 (Fig. 5) to correct the CCB alignment.

### 5.b.3. Cavity alignment

The cavity output beam is separated into its polarization components that are detected by photodetectors Pd1 (CCB) and Pd2 (CAB), see below section 5.b.7. These signals are monitored with the help of an oscilloscope. A correct alignment of the Fabry-Perot cavity should provide resonance peaks with high finesse, which is essential for the experiment.

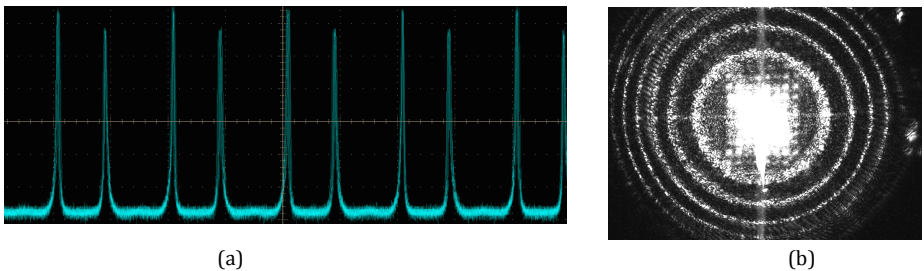

(a) (b)

Fig. 9. Checking of alignment of a Fabry-Perot cavity. In (a) we show peaks of resonance of the cavity with high Finesse. In (b) different longitudinal modes of the cavity are seen in the far field.

As the two mirrors must be perfectly parallel, one must proceed by slightly moving (alternatively) each of the mirrors, always checking the quality of the resonances appearing in the oscilloscope, Fig. 9(a). This is done with the help of a suitable electric modulation that is applied to one of the piezomirrors (PZM2 for CCB or PZM1 for CAB, see Fig. 5). One adjusts the time scale of the oscilloscope to that of the electric signal driving and checks the improvement of the resonance peak. One must also look at the far field and check the presence of well-defined longitudinal modes as shown in Fig. 9(b).

### 5.b.4. Stabilization and control mechanism

The cavity changes its length in some µm because of thermal and mechanical fluctuations, which occur in a slow time scale (in the order of seconds). Then one applies a reference electric signal ($\omega$ = 500 Hz) to one of the piezomirror, PZM2, which modulates the CCB cavity length. This cavity output detected with photodetector Pd2 is sent to a lock-in amplifier together with the reference signal. The lock-in amplifier compares both signals, determines their relative phase, and applies a voltage to the cavity end–mirror PZM3 appropriate for canceling the detected difference. In this way cavity length fluctuations are compensated, even if with some (irrelevant) delay. This procedure corrects the fluctuations of both cavities, CAB and CCB, as their optical lengths are nearly equal (the only difference is the presence of the nonlinear crystal in CAB). However we have also the possibility of acting on PZM1, within the CAB cavity, which is used in the alignments stage for controllably improving the cavity resonances.

An additional continuous drive could be performed over PZM3 to control the cavity detuning as explained above.

### 5.b.5. Pumping the cavity

We mentioned above that the crystal c-axis must be properly oriented with respect to the pumping and the signal beams. Once the cavity has been aligned, the crystal is finally inserted within it. The crystal must be oriented as shown above in Fig. 3. According with the angles shown in Fig. 2. When these values are optimized, one observes *beam-fanning*, see Fig. 10, which appears as a cone of emission that must occur along the cavity axis if everything is properly done. The detection of the near and far fields (see below) will reveal the growing of the light intensity as well as the formation of moving patterns in the near field. Then the active cavity must be slightly realigned because of the changes introduced by the nonlinear crystal in the optical length, this realignment being made by moving with care PZM1, Fig. 5, as already mentioned.

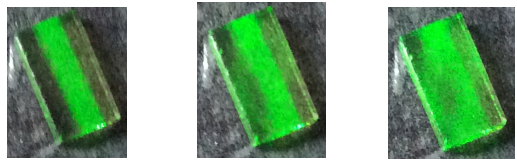

Fig.10. Energy transfer (beam fanning) in a non- degenerate four wave mixing configuration. The three images capture instants separated by 10 sec.

### 5.b.6. Use of additional beams

Up to now, we have considered all the elements necessary for the normal operation of the cavity and in principle nothing more is needed apart from the detection and observation devices that we describe in the following subsection.

But depending on the experiment, additional laser beams can be necessary. In the experiment we are describing an additional rocking beam must be fed into de cavity. The rocking beam is a laser beam of the same frequency as the pumping beam that is amplitude modulated and injected in the cavity. This beam is amplitude modulated before injection by reflecting it on a PZM that oscillates back and forth exactly $\lambda/2$.

This beam must be intense enough for rocking be efficient so it is convenient to have some degree of control on its intensity [20]. In our case, a piezomirror is used to modulate the amplitude, see Fig. 7 by applying an electric AC modulation (frequency 1 to 10Hz).

Another beam that is interesting to have for this experiment is that for writing domains walls in the cavity. The idea is the following: if one injects a laser beam that is tilted with respect to the cavity axis, one is forcing a phase gradient (along the horizontal direction) that depends on the angle the beam forms with the cavity axis. If this gradient covers $2\pi$ one is forcing a phase change in the field so that it will be positive on one side of the crystal and negative on the other. This is the way we inject domain walls in the cavity. Of course, more than one domain wall can be injected by increasing the angle, and consequently the imprinted phase gradient [26].

### 5.b.7. Detection and observation

Our observation requirements include the detection of both the near and far fields, as well as the extraction of some signal for stabilization purposes (lock-in amplifier). Two CCD cameras (Basler scA780-54fm FireWire) are configured to record the far and near fields, for which appropriate image systems must be used. The far field observation is easier to configure because it only needs a lens and the CCD, which is positioned in the focal plane of that lens. The other CCD camera images the near field in the mirror plane closer to the PRO.

The positioning and alignment of the detection mechanism must be done before positioning the intracavity telescopes as already stated. The procedure is the following: the cavity output beam is divided into three beams with the help of beam splitters BS1 and BS2, see Fig. 11. The beam reflected by BS1 is used for stabilization, the one reflected by BS2 is used for the observation of the far field, and the one transmitted by BS2 is used for the observation of the near field

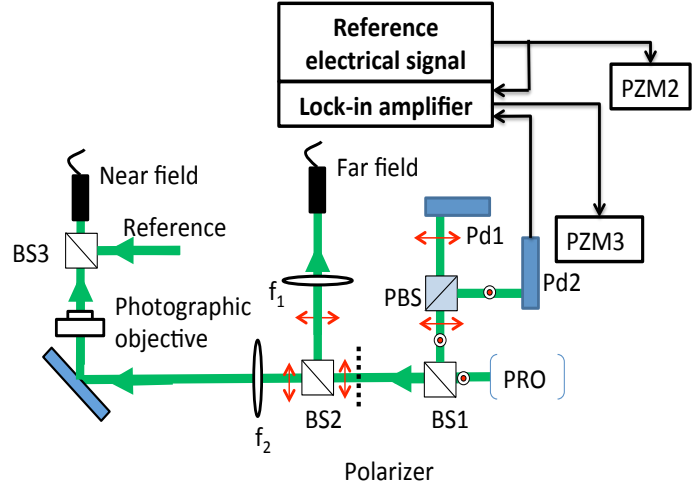

Fig.11. Observation and detection scheme.

The first beam is sent to a PBS that separates the CCB from the CAB contributions to the output, and are detected and sent to the lock-in system. The second beam (reflection in BS2) is sent to a CCD camera through a lens (focal length f= 10 cm in our case), and the transmitted beam is sent to another CCD through a photographic objective (smc-PENTAX-A A/35-75 mm F/3.5- 4.5) and a lens (f= 5 cm). In order to remove the CCB contribution to the cavity output, a linear polarizer is used. Finally, in order to record interferograms of the near field (which is necessary to obtain phase information), we place a beam-splitter BS3 before the second CCD, and inject the reference as indicated in Fig. 11. The detected interferogram is later processed for extracting the amplitude and phase of the field. (See [27–29] for details about how to do that). The scheme of the complete setup is shown in Fig. 12.

## 6. Some experimental results

Even if the purpose of the present article is not to present specific research, here we describe some of the results that we have obtained in the past. Concretely, we are going to present results concerning the demonstration of temporal rocking in order to get a clearer idea of what can be observed in such experiments, and how the recorded patterns must be processed in order to get clear results.

As explained above, a singly–pumped linear cavity PRO is a phase invariant system; hence the system displays vortices, localized structures in which the field intensity has a zero around which the field phase accumulates $2m\pi$ rad with $m$ an integer that is the charge of the vortex.

In Fig. 13(a) we represent the detected intensity distribution in the interferogram of the near field where the presence of several vortices can be appreciated. However, in order to fully characterize these dark points as vortices we must obtain the field phase distribution.

In order to reconstruct the phase and the amplitude map we firstly propagate the interferogram from the object plane to the Fourier plane by using a Fourier transform (FT), which is shown in Fig. 13(b). Then we apply a band-pass filter to one of the images (the upper one is the real image), so that the zero-th order and the other (so called virtual) image are removed, Fig. 13(c), and center it, Fig. 13(d).

The resulting interferogram is again propagated to the image plane, and the amplitude and phase are generated, which are shown in Figs. 13(e) and 13(f). It can be appreciated that in a circuit surrounding one of the dark points in the intensity distribution, the phase accumulates $2\pi$, hence they are vortices. The image can still be improved by numerically adding a spherical wave with variable curvature. This is done with the purpose of compensating the fact that the interferogram has not been taken at the focal plane of the telescopes, and adding a wavefront is similar to propagating the interferogram to the right plane.

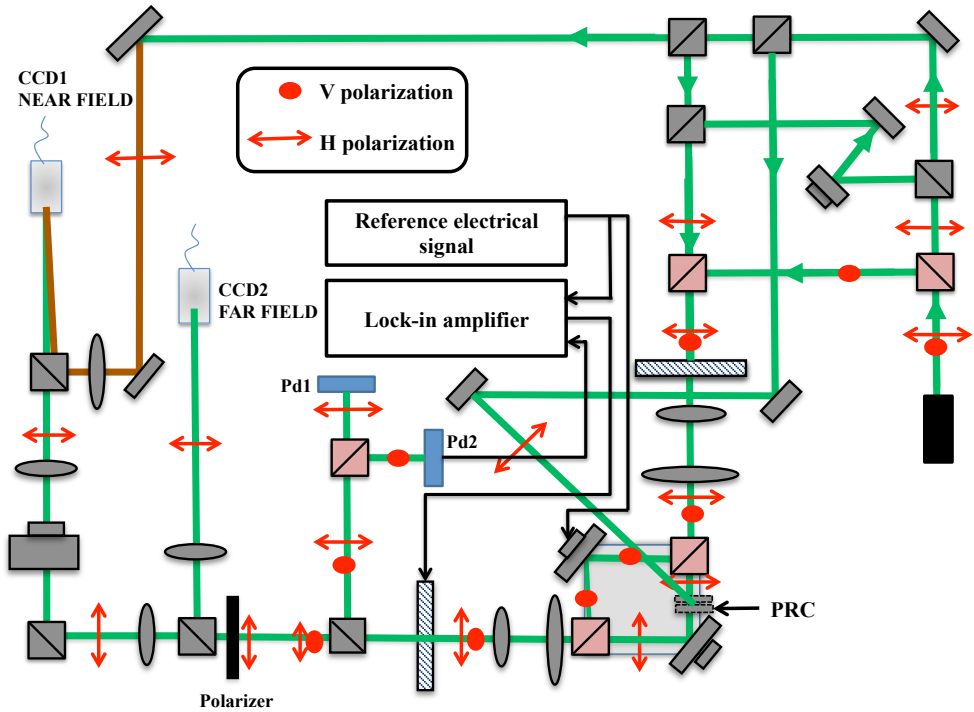

Fig.12. Complete scheme of PRO.

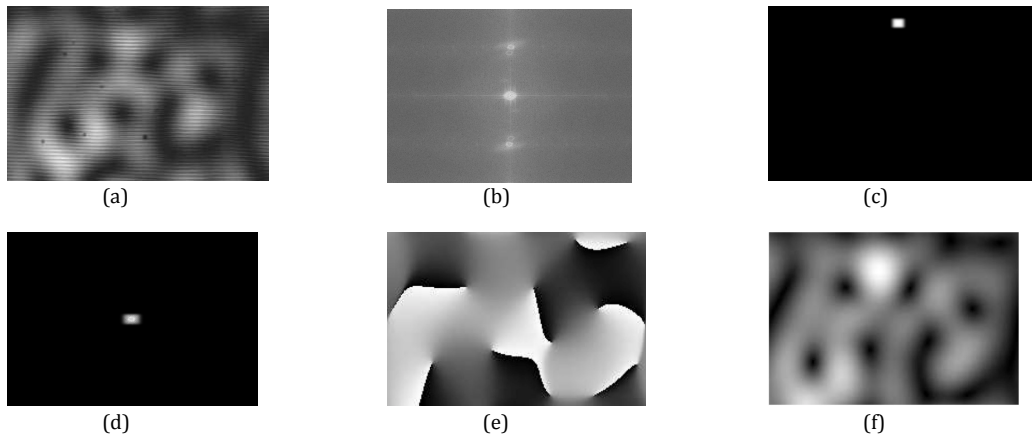

(a)             (b)             (c)

(d)             (e)             (f)

FIG.13. Vortices as seen in the near field. In (a) we show the interferogram, and in (b) we show its Fourier transform where the real image (top image) is selected with a band-pass filter (c) removing the other two images, and later is centered (d). In (e) and (f), an inverse FT is applied. Image (e) shows the phase map reconstruction, while image (e) shows the amplitude map reconstruction.

Now comes the second part of the experiment: converting the system into phase-bistable by injecting the rocking beam. This beam must be intense enough and the amplitude modulation must have a frequency around 10 Hz. The result is shown in Fig. 14 where we represent the result of the processing of the detected interferogram. The appearance of the output is very different to that of Fig. 13, and the presence of domain walls is very clear, see the phase distribution.

The results we have presented here were originally published ten years ago in [32], but our experimental research concerning other rocking techniques, spatial rocking, can be found in [23,25]. At present, we are implementing other rocking schemes [24], as well as investigating the effect of rocking on phase bistable systems.

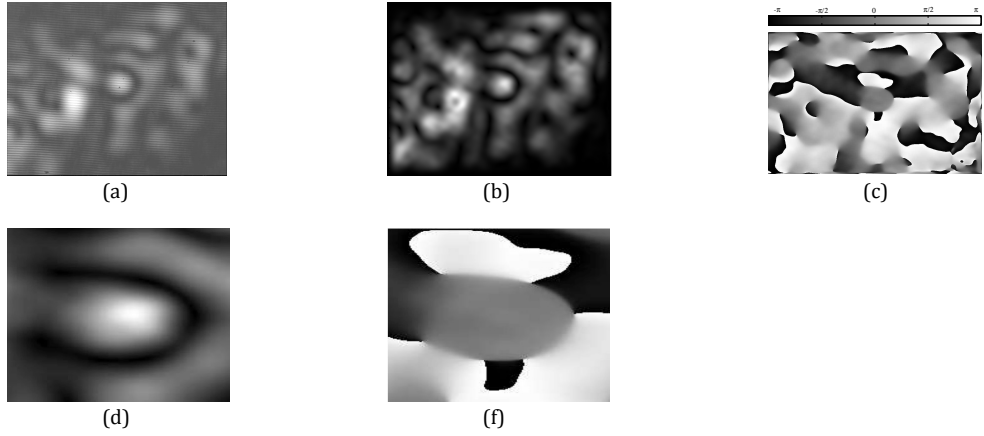

FIG.14. Domain walls as seen in the near field after injecting the rocking beam. (a) Interferogram of the phase domain, see that it is completely different from that shown Fig. 13(a). In (b) and (c) we show the amplitude and phase distributions extracted from (a). In (d) and (e) a magnification is shown in order to better appreciate the details. Inside of the phase domain one sees a homogeneous 0 phase value, whereas outside the phase value changes to +π.

## 7. Conclusions

After a suitable introduction to the photorefractive effect and to the main trends of nonlinear pattern formation in nonlinear optical cavities, we have explained how to build a photorefractive oscillator with large Fresnel number. We have detailed the design and assembling procedure of the different parts of the cavity (the double cavity device, the use of intracavity telescopes, the detection mechanism, etc.), so that the article is reasonably self–contained.

We have also described a special experiment in which one first observes the patterns displayed by the cavity (vortices), and then modifies the output by injecting an amplitude modulated signal (rocking mechanism), which transmutes the vortices into domain walls.

More results concerning temporal rocking can be found in [30–32], as well as more details on the nonlinear dynamics of the system. In fact, in the past the nonlinear dynamics of photorefractive oscillators have been studied by different research groups, and we refer the reader to some of the original references (see [33-53]).

This work has been supported by the Spanish Government (Ministerio de Economía y Competitividad) and the European Union FEDER, through projects FIS2011-26960, FIS2011-29734-C02-11, and FIS2014-60715-P.